# Risk Analysis of Unmanned Aerial System Operations in Urban Airspace Considering Spatiotemporal Population Dynamics *

Soohwan Oh, Yoonjin Yoon, and Seyun Kim, *Member, IEEE*

*Abstract*—This study aims to estimate the fatality risk of Unmanned Aerial System (UAS) operations from a population perspective using high-resolution de facto population data. In doing so, it provides more practical risk values compared to the risk values derived from the residential population data. We then set restricted airspace using the risk values and the acceptable level of safety. We regard the restricted airspace as airspace being blocked by a dynamic obstacle. Scenario analysis on the study area in Seoul, South Korea presents a richer set of results for time-dependent differences in restricted airspace. Especially during the daytime, the restricted airspace is clustered around commercial and business areas. The difference between restricting airspace based on residential population-derived risk and restricting airspace based on de facto population-derived risk is also observed. The findings confirm the importance of accurately taking into account population density when assessing and mitigating the risk of UAS operations. Sensitivity analysis also reveals the importance of an accurate estimate of population density when estimating the risk.

## I. INTRODUCTION

Market service demands of Unmanned Aerial Systems (UAS) are expected to grow with a wide range of applications such as medical aid, food delivery, and package delivery [1], [2]. The expected UAS demand growth has raised safety concerns associated with their use [3]. One of the foremost concerns is the risk of collision between Unmanned Aircraft (UA) and a non-involved person [4]. In order to alleviate these concerns, several countries restrict UAS operations in populated areas to minimize the risk to people on the ground [5]. Such flight restrictions may be issued frequently in urban areas since urban areas are expected to be densely populated. Furthermore, populated areas can appear from time to time and from place to place. This makes it difficult to determine the extent and duration for which reasonable containment boundaries should be defined for minimizing the anticipated operational risk.

In order to determine the reasonable extent and duration of containment boundaries, several researchers have attempted to evaluate the fatality risk to people with various impact area calculation methods and a fixed set of flight failure and fatality probability [4], [6], [7]. However, in most cases, the number of people exposed to an aircraft ground crash is derived based on residential population density. The limitation of this approach is that the residential population does not account for actual people who are exposed to UAS operational risk. Taking a simple example of business districts or metro station neighborhoods during the daytime, the residential population is likely to be minimal whereas actual people exposed to UAS operation can be maximal.

In this regard, Di Donato and Atkins [8] utilized mobile phone activity data together with residential population data to capture active population in a unit cell. Unlike residential population data, mobile phone activity data provide a daytime population that varies significantly depending on the time of the day. Using the mobile phone activity data, they were able to capture anomalous gatherings of people around sports stadiums and alter UA's path to avoid those areas for an emergency landing. One of the limitations is that it does not take into account the fatality risk induced by UAS operations.

The main objective of this study is to estimate more practical risk values of UAS operations by considering spatiotemporal variations of population movement. Using high-resolution de facto population data, we measure the fatality risk of UAS operations for each time of the day. In doing so, it provides more realistic fatality risk values compared to the fatality risk values derived from the residential population data. We then examine the change in airspace operability in terms of shape and volume during a day, by taking spatiotemporal dynamics of population movement into account. We also perform sensitivity analysis to investigate the effect of population density on risk values for multiple parameter value combinations.

The rest of the paper is organized as follows. In Section 2, the relevant literature is summarized. Section 3 describes the details of the risk assessment and data used in this study. In Section 4, the scenario analysis results for the actual urban environment are presented and discussed. Section 5 provides sensitivity analysis results. Finally, conclusions are drawn in Section 6.

## II. RELATED STUDIES

In recent studies, the risk of UAS operations to ground population has assessed in terms of the probability of causing casualties per flight hour [4], [6], [7], [9]. Such probability consists of the likelihood of an event and its consequences [3]. The likelihood can be assessed by considering 'probability of accident' and 'probability of fatality'. The consequence can be assessed by considering 'sheltering effect' and 'population density'. A detailed description of how each element was considered in the relevant literature is as follows.

### A. Probability of accident

A UA can fall into the ground for several causes including malfunctions and collisions with aircraft, but it is difficult to

S. Oh, and Y. Yoon are with Korea Advanced Institute of Science and Technology, Daejeon 34141, South Korea (e-mail: suhwan@kaist.ac.kr, yoonjin@kaist.ac.kr).

S. Kim is with Korea Advanced Institute of Science and Technology, Daejeon 34141, South Korea (corresponding author, e-mail: whataud@kaist.ac.kr).

estimate the probability of an accident considering various causes. One of the reasons for the difficulty is that there are very few real cases of UAS operations. This led most related studies to assume a situation in which a UA is malfunctioning for some reason, regardless of the event type [10]-[12].

The probability of an accident due to malfunction can be obtained from the manufacturer of UA but has not been made available to the public. This situation led relevant studies to assume the probability based on expert opinions. Ford and McEntee [10] assumed that the probability of a catastrophic accident is $10^{-5}$ per flight hour, and the probability of a hazardous accident is $10^{-4}$ per flight hour. A catastrophic accident refers to a situation in which a UA cannot be controlled in the event of a crash, and a dangerous accident refers to a situation in which a UA can be controlled to some extent in the event of a crash. Stevenson et al. [11] assumed different values for the probability of an accident depending on the areas in which UA flies. The probability was assumed to $10^{-5}$ per flight hour in suburban areas and assumed to $10^{-4}$ per flight hour in urban areas. Unlike aforementioned studies that assumed the probability of an accident based on expert opinions, Petritoli et al. [12] estimated the probability based on their actual data. Although the data used to estimate the probability was not disclosed, it is meaningful in that the probability was estimated by classifying malfunctions that could cause a UA fall based on actual data.

*B. Probability of fatality*

When a collision between a UA and a non-involved person occurs, the kinetic energy (KE) is transferred from the UA to the person. The estimation of KE is difficult as it depends on a variety of factors (e.g. materials of UA, initial altitude of UA, angle of impact, etc.). This situation leads several studies to select the KE based on a simple assumption [13]-[18]. Based on expert interviews, Clothier et al. [13] assumed that the probability of fatality is 100%. Melnyk et al. [14] and Burke et al. [15] assumed that the fatality rate is 100% when the KE is more than 80J and 92J, respectively. The values are selected based on historical accident data and literature data. Aalmoes et al. [16] also selected a constant value of fatality rate based on historical accident and literature data, but the authors adopted a value of 17%. Lum et al. [17] adopted the probability of fatality according to a UA's type. The probability of fatality was set to 100% for Predator B weighing more than 300 kg, and 50% for ScanEagle weighing 20 kg. In contrast, Koh et al. [18] calculated the impact energy onto a dummy head for UA falls from various heights through extensive simulations and experiments. They then converted impact energy into the Abbreviated Injury Scale (AIS) for classifying injury levels.

*C. Sheltering effect*

The sheltering effect is a factor that determines the level of shelter provided to people in an area by the presence of buildings or trees. Regarding the sheltering effect, some studies have attempted to take this into account [7], [11], [14], [19]. Weibel and Hansman [19] estimated the sheltering effect based on the KE of the aircraft in cruise. The sheltering effect was set to 5% for micro UA weighing 0.14 pounds and 10% for mini UA weighing 9.6 pounds. Melnyk et al. [14] estimated the sheltering effect not only from the KE but also from root material absorption data. When the KE was less than 2,700J, residential buildings were considered to protect people on the ground, and when the KE was less than 13,500J, commercial buildings were considered to protect people on the ground. Stevenson et al. [11] assumed that the sheltering effect was determined based on terrain. The authors set the value of 75% for urban areas and the value of 25% for wilderness terrain. Recently, Blom et al. [7] assumed that 10% of persons are unprotected in an urban area.

*D. Population density*

Most studies have adopted census data to estimate population density [10], [13], [15], [20]. Clothier et al. [13] calculated the population density using census data, but the spatial resolution of the population density is set according to the size of the census area, so the spatial resolution is low. Burke et al. [15] also use census data to account for population density, but not residential population density. They classified the districts as unpopulated, sparsely populated, densely populated, and open-air assembly, and assumed the same population density values for each category. Ford and McEntee [10] calculated population density based on building density assuming that population density was related to building density. Lum and Waggoner [20] assumed that population density varies with regional characteristics. They set the value of population density as 5 per building in town areas, 20 per square km in the field, and 10 per square km in the forest. Based on this assumption, population densities were estimated using census data and satellite imagery. However, the use of census data has limitations in that spatial resolution is low and temporal information is insufficient.

As summarized in TABLE I, most of the relevant studies have assessed the risk to people posed by UAS operations using a fixed set of flight failure, fatality probability, and sheltering effect. With respect to population density, the number of people at risk of UAS operations was derived based on the residential population density. We question whether the residential population can account for actual people at UAS risk during the daytime. The residential population in business districts is likely to be minimal whereas actual people exposed to UAS operation can be maximal during the daytime. Thus, it cannot fully explain human activity in the daytime. Taking into account that populated regions can appear from time to time and from place to place, it is necessary to utilize de facto population data to the fullest extent and establish a reasonable containment boundary that minimizes anticipated operational risks. In order to address such problems, this study attempts to estimate reasonable fatality risk values of UAS operations by considering spatiotemporal variations of population movement.

TABLE I.     SUMMARY OF LITERATURE ON THE RISK OF UNMANNED AERIAL SYSTEM OPERATIONS TO GROUND POPULATION

| Literature | Probability of accident | Probability of fatality | Sheltering effect | Population density |
|---|---|---|---|---|
| Primatesta et al. [6] | Constant failure rate | KE-based probability | Terrian-based probability | Residential populaiton |
| Blom et al. [7] | Constant failure rate | KE-based probability | Location-based probability | Residential populaiton |
| Clotheir et al. [13] | Constant failure rate | 100% | - | Residential populaiton |
| Melnyk et al. [14] | Constant failure rate | KE-based probability | Material-based probability | Estimated populaiton |
| Burke et al. [15] | Constant failure rate | KE-based probability | KE-based probability | Residential populaiton |
| Weibel and Hansman [19] | Constant failure rate | KE-based probability | Type-based probability | Residential populaiton |
| Lum and Waggoner [20] | Constant failure rate | Mission-based probability | - | Estimated populaiton |

### III. METHODOLOGY

The ultimate goal of risk assessment is to minimize expected operational risks through mitigation actions [9], [11], [14]. One of the mitigation actions to address potential risks of UA flying over people is to set high-risk regions as restricted areas, making airspace inoperable for a specific time [5], [8]. In UAS traffic management airspace, such restricted airspace can be regarded as a dynamic obstacle, and the formation of such obstacles presents a significant impact on the operability of airspace [21]-[23]. In order to establish a reasonable containment boundary of restricted airspace, in this study, we first define and measure the risk of UAS operations to people on the ground as *population risk*. We then set restricted airspace using *population risk* with the acceptable level of safety, and regard it as airspace being blocked by a dynamic obstacle.

Specifically, a dynamic obstacle represents a stationary containment volume that is blocked over critical areas for a specific time. For example, the airspace above stadiums and densely populated areas can become temporarily restricted due to *population risk* and thus be regarded as dynamic obstacles in the air.

*A. Data description*

In this study, de facto population data are collected from Seoul Metropolitan Government Big Data Campus [24]. The data contain the number of people who are physically present in the neighborhood district in Seoul during each one-hour period. Unlike usual resident population counts, de facto population counts include daytime, workplace, and visitor populations and hence can serve as a proxy to analyze people's movements at 1-hour intervals.

The height and shape information of buildings is collected from Ministry of Land, Infrastructure and Transport (MOLIT) of Republic of Korea [25].

*B. Population risk assessment*

As aforementioned in Section 2, the risk of UAS operations to people on the ground is estimated in terms of the probability to cause a fatality per flight hour, which consists of the likelihood of an event and its consequences.

Based on this approach, in this study, we propose a generic metric of *population risk* to measure the risk posed to non-involved people by UAS operations. *Population risk* $r_i^t(d, h)$ is defined as "the probability to cause a fatality per flight hour at location $i$ and time $t$ with respect to a crash of drone weighing $d$ falling from altitude $h$". It satisfies: $r_i^t(d, h) = n_i^t(1 - S_i)P_{fatal}(d,h)P_{failure}(d)$ where $n_i^t$ is the number of people at location $i$ at time $t$, $S_i$ is the percentage of area sheltered by building at location $i$, $P_{fatal}(d,h)$ is the probability of fatality that a crash of drone weighing $d$ falling from altitude $h$ is fatal for an unprotected average person, and $P_{failure}(d)$ is the probability that drone weighing $d$ loses control with the uncontrolled descent with crash on ground.

### IV. SCENARIO ANALYSIS

Population risk can be measured based on various scenarios of UAS operations. Each scenario requires appropriate settings, including UA specifications, mission altitude, failure probability, and fatality probability. In this section, we demonstrate the population risk assessment methodology using a hypothetical scenario of UAS operation in the city of Seoul.

We select the Jung district in the city of Seoul as our case study. Fig. 1 shows the study area, which is one of the major commercial areas with a densely built environment in the entire city.

*A. Scenario Generation*

The hypothetical scenario is based on a multi-copter UA with a weight of 1.1kg flying at an altitude of 30m. The detailed description of scenario settings is as follows.

Regarding the probability of an accident, we assumed that there would be 30.23 system failures in 1 million flight hours over a location y. Petritoli et al. [12] presented the system failure rate of UAS operations per flight hour based on their actual data. With regard to the probability of fatality, it is determined based on the KE. When a collision between UA and a non-involved person occurs, the KE is transferred from the UA to the person. Although the KE derived from a collision of UA depends on a large variety of factors (e.g. materials of UA, initial altitude of UA, angle of impact, etc.), several studies had attempted to estimate impact energy to the person. In this study, we adopted the results of one of the notable studies, Koh et al. [18]. Koh et al. [18] calculated the impact energy from various heights onto a dummy human head through UAV free drop modeling. They then converted the energy into AIS for classifying injury levels. When a UA weighing 1.1kg falls from 30m, the fatality rate is 8-10%. For the sheltering effect, we calculated the percentage of the sheltered area within location $y$ using building data. Finally, we used de facto population data for the population density of location $y$ at time $t$.

With these calculated risk values, we then classified each district to identify restricted airspaces. Each district's airspace is regarded as "restricted" when the risk is over the threshold value of $10^{-7}$, which is the airworthiness certification standard for unmanned aerial vehicles to prevent catastrophic accidents [26].

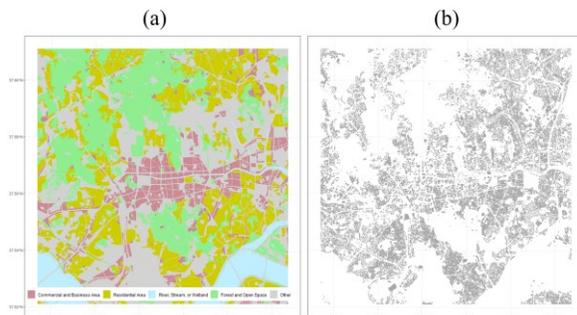

Figure 1. (a) Land use classification; (b) static obstacles in the study area

*B. Empirical Results and Discussions*

In this section, risk maps and restricted airspace maps are illustrated to explore the risk trends over time and geographic location for each spatial unit. The airspace is restricted when population risk exceeds an acceptable level. The acceptable level of safety adopted in this study is $10^{-7}$.

Fig. 2(a) shows a risk map when the risk is derived from the residential population. High-risk regions are observed scattered across several areas. Accordingly, the corresponding airspaces are restricted which occupied 4.84% in the region of interest as shown in Fig. 2(b). The restricted airspaces appeared mainly in residential areas. In Fig. 2(c) and Fig. 2(d), a risk map and a restricted airspace map are illustrated, where the risk is derived from the de facto population data. During 03:00 to 04:00, high-risk regions are observed scattered across several areas, which restrict the corresponding airspaces.

As we expect for the risk derived from the nighttime de facto population, the restricted airspaces are similar to the result derived from the residential population. During the nighttime, people are mostly concentrated in residential areas.

On the other hand, during 13:00 to 14:00, additional high-risk regions and restricted airspaces are observed as illustrated in Fig. 2(e) and Fig. 2(f). It is noticeable that the spatial distribution of the restricted airspaces is clustered around the center compared to other risk maps. This result seems to be due to the characteristics of this area, which is one of the major commercial areas. The population density in commercial areas tends to be higher during the daytime than during the nighttime. This result also reveals that high-risk regions may vary by time of the day.

In addition, when the risk is derived from the de-facto population during 03:00 to 04:00 and 13:00 to 14:00, the restricted airspace accounts for 6.59% and 8.91%, respectively. Compared with the results of the restricted airspace based on the residential population, the results of restricted airspace during 03:00 to 04:00 and 13:00 to 14:00 show a difference of 1.75% and 4.07%, respectively. This shows that there can be a difference between restricting airspace based on the risk derived from the residential population and restricting airspace based on the risk derived from the de facto population depending on the time of the day.

Fig. 3 shows the percentage of restricted airspace during a day. It is calculated as the ratio of restricted airspace area to the total. The restricted airspace ratio was highest during 14:00 to 15:00 and lowest during 09:00 to 10:00. The highest percentage is 9.04% and the lowest is 5.56%. These results seem to be due to the characteristics of commercial districts, where the population density is higher during the day than at night. Changes in restricted airspace with time of the day are observed. During the daytime, the percentage of restricted airspace tends to be high. During the nighttime, it tends to be low. The results suggest that it is necessary to account for time f the day when restricting airspace based on the risks of UAS operations to people. Compared to restricting airspace based on the risk derived from the residential population, there is a big difference during the daytime and a small difference during the daytime and a small difference during the nighttime. These results suggest that restricting airspace based on the risk derived from the residential population may distort the impact of population movements. The proportion of restricted airspace over time does not change abruptly but gradually changes over time. Furthermore, similar values are observed at similar time intervals. This means that a reasonable period setting can be helpful when establishing restricted airspace.

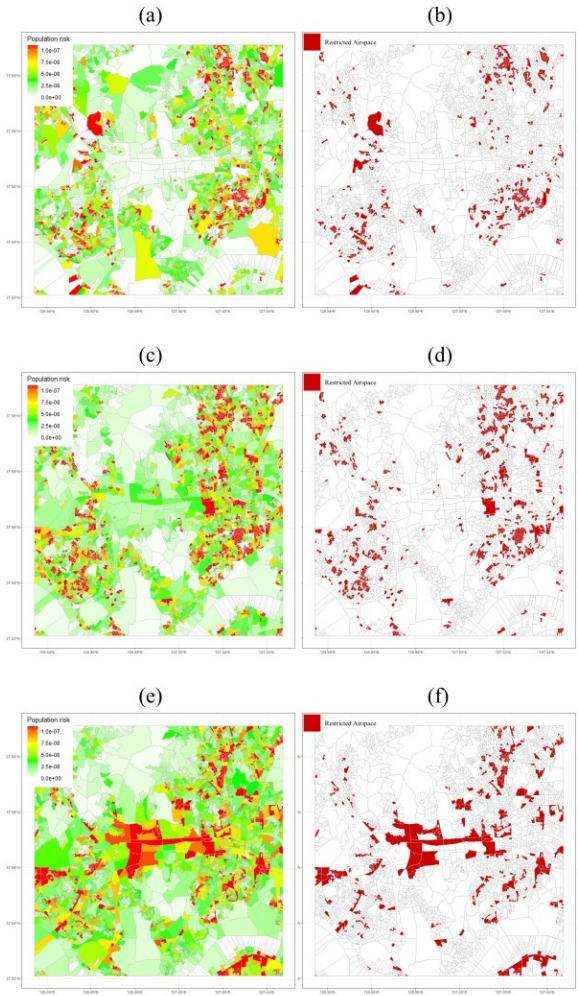

Figure 2. (a) Residential population-derived risk map; (b) residential population-derived restricted airspace map; (c) de facto population-derived risk map (03:00-04:00); (d) de facto population-derived restricted airspace map (03:00-04:00); (e) de facto population-derived risk map (13:00-14:00); (f) de facto population-derived restricted airspace map (13:00-14:00)

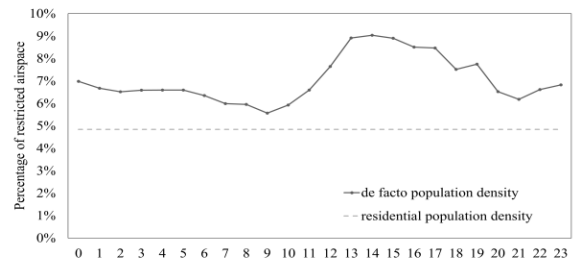

Figure 3. Hourly restricted airspace profile using residential and de facto population density

## V. SENSITIVITY ANALYSIS

As aforementioned in Section 2, $population\ risk$ is estimated based on data collected from existing datasets, published literature, and controlled experiments. This approach generates outcomes, using input parameter estimates ith varying degrees of accuracy due to the availability and quality of the data. However, there may be a lack of reliable empirical data for input parameter values that can drastically alter the model outcome. The large degree of uncertainty around input parameter estimates leads to inaccurate conclusions. Therefore, it is important to measure how variation propagates from the input parameters to outcomes to interpret the importance of each input in determining outcome variability. The sensitivity analysis is one of the methods to capture how an outcome reacts to a change in its input variables, which provides the importance of each input variable in determining outcome variability for combinations of multiple input parameter values [27]-[29]. As a result, it provides insight into which input variables can influence the outcomes and which input variables require a precise estimate to yield accurate outcomes.

In this study, we adopted the Morris method, one of the widely used methods for sensitivity analysis, which provides a measure to quantify the effect of input parameters on the outcome [27]. The method captures outcome variation as one of the sampled points moved to one of the adjacent points. In order to sample a set of points, the input space is discretized by transforming the input parameters into interval variables and dividing each input parameter interval into $p$ levels. Using the discretized input parameters, the magnitude of variation in the model output is calculated according to the following steps until all input parameter values have been explored: (1) sample a set of start values within a defined range of possible values for all input parameters and calculate the corresponding model output, (2) change the value of one variable (all other input parameter values are taken from the remaining start values) and calculate the resulting change in model output compared to the first run, and (3) change the value of another variable (the previous variable remains at the changed value and all other variables remain at their start value) and calculate the resulting change in model output compared to the second run.

This variation due to the defined variation of one input parameter $X$ is elementary effect (EE): $EE_i = (Y(X + e_i \Delta_i) - Y(X))/\Delta_i$ where $e_i$ is a vector of zeros, except for the $i$-th input parameter that equals $\pm 1$. In order to evaluate elementary effects for many combinations $n$, three summary statistics on simulated elementary effects can be calculated and interpreted as sensitivity indicators including the mean $\mu_i$, the absolute mean $\mu_i^*$, and the standard deviation $\sigma_i$. The mean of the elementary effects $\mu_i$ is defined as $\mu_i = 1/n \sum_{j=1}^{k} EE_i^j$. This is interpreted as the average effect of the input variable $i$ on the model output $j$ variation. A high value of $\mu_i$ implies that the input parameter $i$ is an influential variable to the model output, and a low value of $\mu_i$ implies that the input parameter $i$ is a non-influential variable to the model output. The average of the absolute value of the elementary effects is defined as $\mu_i^* = 1/n \sum_{j=1}^{k} |EE_i^j|$. Compared to the mean $\mu_i$, the absolute mean $\mu_i^*$ is a more robust sensitivity indicator ensuring robustness against non-monotonic models. Similar to $\mu_i$, a high value of $\mu_i^*$ implies that the input parameter $i$ is an influential variable to the model output, and a low value of $\mu_i^*$ implies that the input parameter $i$ is a non-influential variable to the model output. However, it is still an average and does not account for non-linear effects. Taking into account non-linear effects between input variables, the standard deviation of the elementary effects $\sigma_i$ is computed as $\sigma_i = \sqrt{1/(n-1) \sum_{j=1}^{k} (EE_i^j - \mu_i)^2}$.

A sensitivity analysis was performed to determine which input variables affected $population\ risk$. The input variables and their parameter values are shown in TABLE II. All potential parameter values varied within reasonable ranges, based on data collected from existing datasets, published literature, and controlled experiments.

The result is shown in Fig. 4. In Fig. 4, the points represent the magnitude of variation in the risk due to one input parameter value change at a time. A higher absolute mean indicates that the input variable has a greater impact on the risk. The input variable of 'population density' has the greatest influence on $population\ risk$. Another group of input variables with slightly lower values includes 'sheltering effect', 'probability of accident', and 'probability of fatality'. Overall, the ranking in Fig. 4. is consistent with findings from previous literature, which identify 'UAS weight', 'population density', 'vehicle failure rate' as influential input variables for the risk of UAS operations [14].

TABLE II. INPUT PARAMETER ESTIMATES USED IN MORRIS METHOD

| Parameter | Data Source | Low estimate | High estimate |
|---|---|---|---|
| Probability of accident (%) | Ford and McEntee [10], Stevenson et al. [11], Petritoli et al. [12], Clothier et al. [13] | $30.23 \times 10^{-4}$ | $10^{-2}$ |
| Probability of fatality (%) | Clothier et al. [13], Melnyk et al. [14], Burke et al. [15], Aalmoes et al. [16], Lum et al. [17], Koh et al. [18] | 0 | 100 |
| Sheltering effect (%) | National Spatial Data Infrastructure Portal [25] | 0 | 100 |
| Population density (people/$m^2$) | Seoul Open Data Plaza [24] | 0 | 0.55841 |

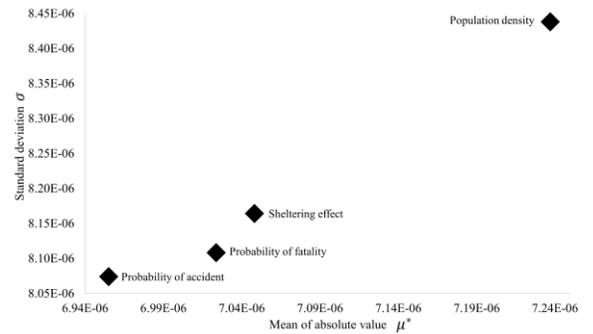

Figure 4. The results of sensitivity analysis by the Morris method

## VI. Conclusion

In this study, we delivered more practical risk values of UAS operations from a population perspective. Using high-resolution de facto population data, we measured the population risk of UAS operations for each time of the day. This provided more practical risk values compared to the risk values derived from the residential population. Furthermore, we examined the change in airspace operability in terms of shape and volume during a day, by restricting the airspace in which the risk value is above the given threshold.

We conducted scenario analysis on the Jung district in Seoul, South Korea, and found that there is a difference in the restricted airspace by time of the day. Especially in the daytime, the restricted airspaces were clustered around the business areas. We also found that there is a difference between restricting airspace based on residential population-derived risk and restricting airspace based on de facto population-derived risk. This finding confirmed the importance of accurately estimating population density by time of the day when assessing and mitigating the population risk of UAS operations.

One of the limitations of this study is that our approach does not take into account the traffic density of UAS operations. Without considering the traffic density of UAS operations, we rather conservatively restrict the airspaces as a whole. Future studies may benefit from the realistic traffic density of UAS operations, for defining more reasonable restricted airspace boundaries.

## Acknowledgment

This work is financially supported by Korea Ministry of Land, Infrastructure and Transport(MOLIT) as 「Innovative Talent Education Program for Smart City」 and is supported by the fourth Stage of Brain Korea FOUR 21 Project in 2021.